\documentclass[ejs]{imsart}

\RequirePackage[OT1]{fontenc}
\RequirePackage{amsthm,amsmath}
\RequirePackage[numbers]{natbib}
\RequirePackage[colorlinks,citecolor=blue,urlcolor=blue]{hyperref}

\pubyear{2014}
\volume{0}
\issue{0}
\firstpage{}
\lastpage{}

\startlocaldefs
\numberwithin{equation}{section}
\theoremstyle{plain}

\newtheorem{Lem}{Lemma}
\newtheorem{Theo}{Theorem}
\newtheorem{Coro}{Corollary}
\endlocaldefs

\begin{document}

\begin{frontmatter}
\title{Uniform-in-bandwidth consistency \\ for nonparametric estimation of divergence measures}
\runtitle{nonparametric estimation of divergence measures}

\begin{aug}
\author{\fnms{Hamza} \snm{Dhaker$^{\dagger,\ddagger}$}\thanksref{}\ead[label=e1]{hamzayahya@hotmail.fr}},
\author{\fnms{{\bf Papa} } \snm{{\ \bf Ngom}$^{\dagger,\ddagger}$}\ead[label=e2]{papa.ngom@ucad.edu.sn}},\\
\author{\fnms{{\bf Pierre}} \snm{{\ \bf Mendy}$^{\ddagger}$}\ead[label=e3]{fmendy$\_$99@yahoo.fr}}
\author{\fnms{{\bf and \ El \ Hadji}}\snm{{\ \bf Deme}$^{\S}$}
\ead[label=e4]{elhadjidemeufrsat@gmail.com}}
\address{{\bf LMA}$^{\dagger}$,\ Universit\'e \ Cheikh \ Anta \ Diop, \ BP\ 5005\ Dakar-Fann, \ S\'en\'egal\\
{\bf LMDAN}$^{\ddagger}$, \ Universit\'e \ Cheikh \ Anta \ Diop, \ BP \ 5005 \ Dakar-Fann, \ S\'en\'egal\\
{\bf LERSTAD}$^{\S}$, {UFR} \ {SAT} \  Universit\'e Gaston \ Berger,\  BP \ 234 \ Saint-Louis,\ S\'en\'egal\\
\printead{e1}
\printead{e2}\\
\printead{e3}
\printead{e4}}
\runauthor{H. Dhaker et al.}

\end{aug}

\begin{abstract}
We propose nonparametric estimation of divergence measures between continuous
distributions. Our approach is based on a plug-in kernel-type estimators of density functions.
We give the uniform in bandwidth consistency for the proposal estimators.
As a consequence, their asymptotic 100\% confidence intervals are also provided.
\end{abstract}


\tableofcontents
\end{frontmatter}

\section{Introduction}

Given samples from two distributions, one fundamental and classical question to ask is: how close are the
two distributions? First, one must specify what it means for two distributions to be close, 
for which many different measures quantifying the degree of these distributions have been studied in the past.
They are frequently called distance measures, although some of them are not strictly
metrics. 
The divergence measures play an important role in statistical theory, especially in large theories of estimation and
testing.
They have been applied to different areas, such
as medical image registration (\cite{r25}),
classification and retrieval. 
In machine learning, it is often convenient to view training data as a set of distributions and use divergence measuires to estimate dissimilarity
between examples. This idea has been used in neuroscience, where the neural response pattern of an individual is modeled as a distribution, 
and divergence meaures is used to compare responses across subjects (see, e.g \cite{r21}). 
Later many papers have appeared in the literature,
where divergence or entropy type measures of information have been used in testing statistical hypotheses.
For more examples and other possible applications of divergence measures, see the extended technical report (\cite{r27,r28}).
For these applications and others, it is crucial to accurately estimate divergences.\\
The class of divergence measures is large; it includes the R\'enyi-$\alpha$ (\cite{r29,r30}), Tsallis-$\alpha$ (\cite{r34}), Kullback-Leibler (KL), 
Hellinger, Bhattacharyya, Euclidean divergences, etc.
These divergence measures can be related to the Csisz\'ar-$f$ divergence (\cite{r5}). 
The  Kullback-Leibler, Hellinger and Bhattacharyya are special cases of
R\'enyi-$\alpha$ and Tsallis-$\alpha$ divergences. 
But the Kullback Leibler one is the most popiular of these divergence measures. 
\vskip1ex  
\noindent

In the nonparametric setting, a number of authors have proposed various estimators which are provably consistent. Krishnamurthy and Kandasamy
\cite{r22} used an initial plug-in estimator by estimates of the higher
order terms in the von Mises expansion of the divergence functional.
In their frameworks, they proposed tree estimators for 
R\'enyi-$\alpha$, Tsallis-$\alpha$, and Euclidean divergences between two
continuous distributions and establised the rates of convergence of these estimators.
\vskip1ex  
\noindent
The main purpose of this paper is to analyze estimators for divergence measures between two
continuous distributions. 
Our approach is similar on those of Krishnamurthy and Kandasamy \cite{r22} and is based on plug-in estimation scheme:
first, apply a consistent density estimator for the underlying
densities, and then plug them into the desired formulas.
Unlike of their frameworks, we study the uniform bandwidth consistent estimators of these divergences.
We introduce a method to establish consistency 
of kernel-type estimators divergences between two
continuous distributions when the bandwidthh is allowed to range in a small
interval which may decrease in length with the sample size. Our results will be immediately applicable to proving 
uniform bandwidth consistency for nomparametric estimation of divergenge measures.
\vskip1ex  
\noindent
The rest of this paper is organized as follows: in Section 2, we introduce divergence measures and we construct their nonparametric estimators.
In Section 3, we study the unfiform bandwidth consistency of the proposal estimators. Section 4 is devoted on the proofs.

\section{Divergence Measures and Estimation}
Let us begin by standardizing notation and presenting some basic definitions. We will be concerned with two densities, $f\ ,\ g$ : 
$\mathrm{R}^d 
\mapsto [0,1]$ 
where $d\geq 1$ denotes the dimension.
The divergence measures of interest are R\'enyi-$\alpha$, Tsallis-$\alpha$, Kullback-Leibler, 
Hellinger, Bhattacharyya are  defined respectivelly as follows

\begin{eqnarray}
{ D}_{\alpha}^R (f,g)&
 = &\frac{1}{\alpha-1}\log \int_{\mathrm{R}^d}f^\alpha(x)g^{1-\alpha}(x)dx,\ \ \ 
\alpha \in \mathrm{R}\setminus \lbrace 1\rbrace \\ 
{\cal D}_{\alpha}^T(f,g)&=&\frac{1}{\alpha-1}\left(\int_{\mathrm{R}^d}f^\alpha(x)g^{1-\alpha}(x)dx-1\right),\ \ \ \alpha \in \mathrm{R}\setminus \lbrace 1\rbrace\\
{\cal D}^{KL}(f,g)&=&\int_{\mathrm{R}^d}f(x)\log\frac{f(x)}{g(x)}dx,\\
{\cal D}^H(f,g)&=&1-\int_{\mathrm{R}^d}f^{1/2}(x)g^{1/2}(x)dx,\\
{\cal D}^B(f,g)&=&-\log\int_{\mathrm{R}^d}f^{1/2}(x)g^{1/2}(x)dx,
\end{eqnarray}

whenever the integrals in the underlying expressions are meaningful.
These quantities are nonnegative, and they are zero iff $f = g$ almost 
surely ($\it{a.s}$).
These expressions can be used to
measure the distance between two distributions.
Remark that, the divergences ${\cal D}^H(f,g)$ and ${\cal D}^B(f,g)$ are respectively special cases of ${\cal D}_{\alpha}^T(f,g)$ and ${\cal D}_{\alpha}^R(f,g)$.
We easily check that  
$$\displaystyle\lim_{\alpha\longrightarrow1}{\cal D}_{\alpha}^R(f,g)=
{\cal D}^{KL}(f,g).$$
For the following, we focus only on the estimation of ${\cal D}_{\alpha}^T(f,g)$ and ${\cal D}_{\alpha}^R(f,g)$.
The  Kullback-Leibler, Hellinger and Bhattacharyya can be deducing immediately.
\\
We will next provide consistent estimator for the following quantity
\begin{eqnarray}
\label{eqdiv}
 {\cal D}_{\alpha}(f,g)&=& 
 \int_{\mathrm{R}^d}f^\alpha(x)g^{1-\alpha}(x)dx,
\end{eqnarray}
whenever this integral is meaningful.
Plugging it estimates into the appropriate formula immediately leads to consistent estimator for 
the divergence measures ${\cal D}_{\alpha}^R(f,g)$, ${\cal D}_{\alpha}^T(f,g)$.\\
\vskip1ex  
\noindent
Now, assuming that for all the rest of the paper, the density $g$ satisfies : 
$\int_{\mathrm{R}^d} g^{1-\alpha}(x)dx$ 
is finite, this implies that ${\cal D}_{\alpha}(f,g)$ is finite. Next, consider $X_1,...,X_n,\\ n \geq 1$ a sequence of independent and identically distributed $\mathrm{R}^d$-valued
random vectors, with cumulative distribution function $F$ a density function $f(\cdot)$ with respect to Lebesgue measure on
$\mathrm{R}^d$.
We start by giving some notation and conditions that are needed for the
forthcoming sections.
To construct our divergence estimators we define, in a first step, a kernel density
estimator for $f(\cdot)$, and then substituting $f(\cdot)$ by its estimator in the divergence like functional of $f(\cdot)$.
Towards this aim, we introduce a measurable function $K(\cdot)$ fulfilling
the following conditions.
\vskip1ex  

\ (\textbf{K.1}) $K(\cdot)$ is of bounded variation on $\mathrm{R}^d $
\vskip1ex  

\ (\textbf{K.2}) $K(\cdot)$ is right continuous on $\mathrm{R}^d $
\vskip1ex  

\ (\textbf{K.3}) $|| K||_{\infty}=\displaystyle \sup_{x\in \mathrm{R}^d} \mid K(x) \mid  < \infty $
\vskip1ex  

\ (\textbf{K.4}) $\int_{\mathrm{R}^d}K(t)dt=1.$
\vskip1ex  
\noindent
The well known Akaike-Parzen-Rosenblatt (refer to \cite{r1,r23} and \cite{r31}) kernel estimator of $f(\cdot)$ is defined, for any $x \in \mathrm {R}^d$, by
\begin{equation}
\label{kernel}
\widehat{f}_{n,h_{n}}(x)= \frac{1}{nh_{n}^{d}}\sum_{i=1}^{n} K 
\left( \frac{x-X_i}{h_n} \right),
\end{equation}
where $0 < h_n \leq 1$  is the smoothing parameter. 
For notational convenience, we have chosen the same bandwidth sequence for each margin.
Assuming that the density $f$ is contiuous, one obtains a strongly consistent estimator $\widehat{f}_{n,h_n}$ of $f$,
that is, one has with probability $1$, $\widehat{f}_{n,h_n}(x)\longrightarrow f(x)$, $x\in\mathrm{R}^d$. 
There are also results concerning uniform convergence and convergence rates.
For proving such results one usually writes the difference $\widehat{f}_{n,h_n}(x)-f(x)$ as the sum of a probabilistic term
$\widehat{f}_{n,h_n}(x)-\mathrm E \widehat{f}_{n,h_n}(x)$ and a deterministic term $\mathrm E \widehat{f}_{n,h_n}(x)-f(x)$, the so-called bias.
On can refer to \cite{r14,r17,r19} , among other authors.
\vskip1ex  
\noindent
In a second step, given $\widehat{f}_{n,h_n}(\cdot)$, we estimate ${\cal D}_{\alpha}(f,g)$ by setting
\begin{eqnarray}
\label{divestim}
\widehat{{\cal D}_{\alpha}(\widehat{f}_{n,h_n},g)} &=&\int_{A_{n,h_n}}\widehat{f}_{n,h_n}^\alpha(x)g^{1-\alpha}(x)dx,\ \ \ \alpha\neq1
\end{eqnarray}
where $ \textit{A}_{n,h_n}=\lbrace x\in\mathrm{R}^d, \ \  \widehat{f}_{n,h_n}(x) \geq \gamma_n\rbrace$ 
and $\gamma_n\downarrow0$ is a sequence of positive constant.
Thus, using \ref{divestim}, the associated divergences ${\cal D}_{\alpha}^R(f,g)$ and ${\cal D}_{\alpha}^T(f,g)$ can be estimated by:
\begin{eqnarray*}
\widehat{{\cal D}}_{\alpha}^R(\widehat{f}_{n,h_n},g)&=&\frac{1}{\alpha-1}\log \widehat{{\cal D}}_{\alpha}(\widehat{f}_{n,h_n},g),\\
&\;&\\
\widehat{{\cal D}}_{\alpha}^T(\widehat{f}_{n,h_n},g)&=&\frac{1}{\alpha-1}\left( \widehat{{\cal D}}_{\alpha}(\widehat{f}_{n,h_n},g)-1\right).
\end{eqnarray*}
The appraoch use to define the plug-in estimators is also develloped in \cite{r3}
in order to introduce a kernel-type estimators of Shannon's entropy.
The uniform bandwidth of these divergences is related on those of the kernel estimator $\widehat{f}_{n,h_n}(\cdot)$. 
\vskip1ex  
\noindent
The limiting behavior of $\widehat{f}_{n,h_n}(\cdot)$, for appropriate choices of the bandwidth $h_n$,
has been studied by a large number statisticians over many decades. For good sources of references to research literature in
this area along with statistical applications consult \cite{r10,r11,r2} and \cite{r26}. In particular, under our assumptions,
the condition that $h_n\downarrow0$ together with $nh_n\uparrow\infty$ is necessary and sufficient for the convergence in probability of 
$\widehat{f}_{n,h_n}(x)$ towards the limit $f(x)$, independently of $x\in\mathrm{R}^d$ and the density $f(\cdot)$. 
Various uniform consistency results involving the
estimator $\widehat{f}_{n,h_n}(x)$ have been established.
 We refer to \cite{r6,r14,r9} and the references therein.
In the next section, we will use their methods to establish convergence results for the estimates  
$\widehat{{\cal D}}_{\alpha}(\widehat{f}_{n,h_n},g)$ and deduce 
the convergence results of $\widehat{{\cal D}}_{\alpha}^R(\widehat{f}_{n,h_n},g)$ and $\widehat{{\cal D}}_{\alpha}^T(\widehat{f}_{n,h_n},g)$.

\section{Main Results}
We first study the strong consistency of the estimator $\widehat{{\cal D}}_{\alpha}(\widehat{f}_{n,h_n},g)$
defined in (\ref{divestim}).
We shall consider another, but more appropriate and more computationally convenient, centering factor than the
expectation $\mathrm E\widehat{{\cal D}}_{\alpha}(\widehat{f}_{n,h_n},g)$ which is delicate to handle. This is given by
$$ \mathrm{\widehat{E}}\widehat{{\cal D}}_{\alpha}(\widehat{f}_{n,h_n},g):=\int_{A_{n,h_n}}\left(\mathrm{E}\widehat{f}_{n,h_n}(x)\right)^\alpha
g^{1-\alpha}(x)dx.$$
\vskip1ex  
\noindent
\begin{Lem}
\label{lem1}
Let $K(\cdot)$ satisfy (\textbf{K.1-2-3-4}) and let $f(\cdot)$ be a continuous bounded density
. Then, for each pair of sequence $(h_{n}^{'})_{n\geq1}$, $(h_{n}^{''})_{n\geq1}$ such that 
$0< h_{n}^{'} < h_{n} \leq h_{n}^{''}$, together with $h_{n}^{''} \longrightarrow 0 $, 
$\displaystyle nh_{n}^{\prime}/\log(n)\longrightarrow \infty $ as $n \longrightarrow \infty $, 
for any $\alpha\in(0,1),$ one has with probability 1

\begin{equation}
 \displaystyle \sup_{h_{n}^{'}\leq h \leq h_{n}^{''}}\left|\widehat{{\cal D}}_{\alpha}(\widehat{f}_{n,h},g)-\mathrm{\widehat{E}}\widehat{{\cal D}}_{\alpha}(\widehat{f}_{n,h},g)\right| =
O\left(\left(\frac{\log(1/h_n^\prime)\vee \log \log n}{nh_{n}^\prime}\right)^{\alpha/2}\right) \nonumber. 
\end{equation}

\end{Lem}
\vskip1ex  
\noindent
The proof of Lemma \ref{lem1} is postponed until Section \ref{proof}.
\begin{Lem}
\label{lem2}
Let $K(\cdot)$ satisfy (\textbf{3-4}) and let $f(\cdot)$ be a uniformly Lipschitz and continuous density. 
Then, for each pair of sequence $(h_{n}^{'})_{n\geq1}$, $(h_{n}^{''})_{n\geq1}$ such that 
$0< h_{n}^{'} < h_{n} \leq h_{n}^{''}$, together with $h_{n}^{''} \longrightarrow 0 $, 
as $n \longrightarrow \infty $, for any $\alpha\in(0,1),$
we have
\begin{equation}
 \displaystyle\sup_{h_{n}^{'}\leq h \leq h_{n}^{''}}\left|\mathrm{\widehat{E}}\widehat{{\cal D}}_{\alpha}(\widehat{f}_{n,h},g)-
{\cal D}_{\alpha}(f,g)\right|=O\left(\gamma_n^\alpha 
\vee h_n^{\prime \prime {\alpha/d}}\right). \nonumber
\end{equation}
\end{Lem}
\vskip1ex  
\noindent
The proof of Lemma \ref{lem2} is postponed until Section \ref{proof}.
\begin{Theo}
\label{theo1}
Let $K(\cdot)$ satisfy (\textbf{K.1-2-3-4}) 
and let $f(\cdot)$ be a uniformly Lipschitz, bounded and continuous density. Then, for each pair of sequence $(h_{n}^{'})_{n\geq1}$, $(h_{n}^{''})_{n\geq1}$ such that 
$0< h_{n}^{'} < h_{n} \leq h_{n}^{''}$, together with $h_{n}^{''} \longrightarrow 0 $, 
$\displaystyle nh_{n}^{\prime}/\log(n)\longrightarrow \infty $ as $n \longrightarrow \infty $, 
for any $\alpha\in(0,1),$ one has with probability 1
\begin{equation}
 \displaystyle\sup_{h_{n}^{'}\leq h \leq h_{n}^{''}}\left|\widehat{{\cal D}}_{\alpha}(\widehat{f}_{n,h},g)-{\cal D}_\alpha(f,g)\right| =
O\left(\!\!\left(\frac{\log(1/h_n^\prime)\vee \log \log n}{nh_{n}^\prime}\right)^{\alpha/2}\!\!\!\!\!\!\!\vee\gamma_n^\alpha\vee h_n^{\prime\prime {\alpha/d}}\right).\nonumber
 \end{equation}
 This, in turn, implies that
\begin{equation}
\label{amza1}
 \displaystyle \lim _{n\rightarrow \infty }\sup_{h_{n}^{\prime}\leq h\leq h_{n}^{\prime\prime}}
\left|\widehat{{\cal D}}_{\alpha}(\widehat{f}_{n,h},g)-{\cal D}_{\alpha}(f,g)\right|=0 \quad \quad a.s. 
\end{equation}
\end{Theo}
\noindent The proof of Theorem \ref{theo1} is postponed until Section \ref{proof}.                             
\\
The following corollaries handle respectively the uniform deviation of the estimate $\widehat{{\cal D}}_{\alpha}^T(\widehat{f}_{n,h},g)$ and $\widehat{{\cal D}}_{\alpha}^R(\widehat{f}_{n,h},g)$
with respect to ${\cal D}_{\alpha}^T(f,g)$ and ${\cal D}_{\alpha}^R(f,g)$.
\begin{Coro}
\label{coro1}
Assuming that the assumptions of Theorem \ref{theo1} hold. Then, we have
\begin{equation}
 \displaystyle\sup_{h_{n}^{'}\leq h \leq h_{n}^{''}}\left|\widehat{{\cal D}}_{\alpha}^T(\widehat{f}_{n,h},g)-{\cal D}_\alpha^T(f,g)\right| =
O\left(\left(\frac{\log(1/h_n^\prime)\vee \log \log n}{nh_{n}^\prime}\right)^{\alpha/2}\!\!\!\!\!\!\!\vee\gamma_n^\alpha\vee h_n^{\prime\prime {\alpha/d}}\right).\nonumber
 \end{equation}
 This, in turn, implies that
\begin{equation}
\label{amza2}
 \displaystyle \lim _{n\rightarrow \infty }\sup_{h_{n}^{\prime}\leq h\leq h_{n}^{\prime\prime}}
\left|\widehat{{\cal D}}_{\alpha}^T(\widehat{f}_{n,h},g)-{\cal D}_{\alpha}^T(f,g)\right|=0 \quad \quad a.s. 
\end{equation}
\end{Coro}
\vskip1ex  
\noindent
The proof of Corollary \ref{coro1} is postponed until Section \ref{proof}.
\vskip1ex  
\noindent
\begin{Coro}
\label{coro2}
Assuming that the assumptions of Theorem \ref{theo1} hold. Then, we have
\begin{equation}
 \displaystyle\sup_{h_{n}^{'}\leq h \leq h_{n}^{''}}\left|\widehat{{\cal D}}_{\alpha}^R(\widehat{f}_{n,h},g)-{\cal D}_\alpha^R(f,g)\right| =
O\left(\left(\frac{\log(1/h_n^\prime)\vee \log \log n}{nh_{n}^\prime}\right)^{\alpha/2}\!\!\!\!\!\!\!\vee \gamma_n^\alpha\vee h_n^{\prime\prime {\alpha/d}}\right)\nonumber
 \end{equation}
 This, in turn, implies that
\begin{equation}
 \label{amza3}
 \displaystyle \lim _{n\rightarrow \infty }\sup_{h_{n}^{\prime}\leq h\leq h_{n}^{\prime\prime}}
\left|\widehat{{\cal D}}_{\alpha}^R(\widehat{f}_{n,h},g)-{\cal D}_{\alpha}^R(f,g)\right|=0 \quad \quad a.s. 
\end{equation}
\end{Coro}
\vskip1ex  
\noindent
The proof of Corollary \ref{coro2} is postponed until Section \ref{proof}.\\
Note that, the main problem in using the divergence estimates such
as (\ref{divestim}) is to choose properly the smoothing parameter $h_n$.
The result given in (\ref{amza1}), (\ref{amza2}) and (\ref{amza3}) show that any choice of $h$ between $h_n^\prime$ and $h_n^{\prime\prime}$
ensures the consistency of the underlying divergenge estimates. In other word, the fluctuation of the
bandwidth in a small interval do not affect the consistency of the nonparametric
estimator of these divergences.
\vskip1ex  
\noindent
Now, we shall establish another result in a similar direction for a class of
compactly supported densities. We need the following additional conditions.
\vskip1ex
${\bf F.1}$  $f(\cdot)$ has a compact support say $\mathrm{I}$ and is is $s$-time continuously differentiable,
and there exists a constant $0 < M < \infty$ such that
$$\sup_{x\in\mathrm{I}}\left|\frac{\partial^sf(x)}{\partial x_1^{j_1}...\partial x_d^{j_d}}\right|\leq M,\ \ j_1+\cdots+j_d=s.$$
\vskip1ex
\ (\textbf{K.5}) $K(\cdot)$ is of order $s$, i.e., for some constant $\varrho\neq 0,$
$$ \int_{\mathrm{R}^d}u_1^{j_1}...u_d^{j_d} K(u)du=0,\ \ \ j_1,..., j_d\geq 0,\ \ j_1+\cdots+j_d=1,...,s-1,$$
\ \ \ \ \ \ \ \ \ \ \ \ \ and
$$ \int_{\mathrm{R}^d}|u_1^{j_1}...u_d^{j_d} |K(u)du=\varrho,\ \ j_1,...,j_d\geq 0,\ \  j_1+\cdots+j_d=s.$$

\noindent Under {({\bf F.1})} the expression ${\cal D}_{\alpha}(f,g)$ may be written as
follows
\begin{equation}
\label{diverg111}
{\cal D}_{\alpha}(f,g)=\int_{\mathrm{I}}f^\alpha(x)g^{1-\alpha}dx.
\end{equation}
\begin{Theo}
\label{theo2}
Assuming conditions (\textbf{K.1-2-3-4-5}) hold. 
Let $f(\cdot)$ fulfill (\textbf{F.1}).
 Then for each pair of sequences 
$0< h_{n}^{'} < h_{n} \leq h_{n}^{''}$ with $h_{n}^{''} \longrightarrow 0 $, $nh_n^\prime/\log n\longrightarrow \infty$ as $n\longrightarrow\infty$, 
for any $\alpha\in(0,1),$, we have 
\begin{equation}
\displaystyle \limsup_{n\longrightarrow \infty} \displaystyle \sup_{h_{n}^{\prime}\leq h\leq h_{n}^{\prime \prime}} \frac{\sqrt{(nh)^{\alpha}}
\left|\widehat{{\cal D}}_{\alpha}(\widehat{f}_{n,h},g)-{\cal D}_{\alpha}(f,g)\right| }{\sqrt{(\log(1/h)\vee \log \log n)^{\alpha}}}\leq \zeta(\mathrm{I})\int_{\mathrm{R}^d}g^{1-\alpha}(x)dx \ \ 
\hbox{ a.s },\nonumber
\end{equation}  
where
$$\zeta(\mathrm{I})=\sup_{x\in \mathrm{I}}\left\{f(x)\int_{\mathrm{R}^d} K^2(u)du\right\}^{\alpha/2}.$$
\end{Theo}
\vskip1ex  
\noindent
The proof of Theorem \ref{theo2} is postponed until Section \ref{proof}.
\begin{Coro}
 \label{coro3}
Assuming that the assumptions of the Theorem \ref{theo2} hold.
Then,
\begin{equation}
\displaystyle \limsup_{n\longrightarrow \infty} \displaystyle \sup_{h_{n}^{\prime}\leq h\leq h_{n}^{\prime\prime}} \frac{\sqrt{(nh)^{\alpha}}
\left|\widehat{{\cal D}}_{\alpha}^T(\widehat{f}_{n,h},g)-{\cal D}_{\alpha}^T(f,g)\right|
 }{\sqrt{(\log(1/h)\vee \log \log n)^{\alpha}}}\leq \frac{1}{1-\alpha}\zeta(\mathrm{I})\int_{\mathrm{R}^d}g^{1-\alpha}(x)dx \ \ \hbox{ a.s },\nonumber
\end{equation}  
\end{Coro}
\vskip1ex  
\noindent
\begin{Coro}
 \label{coro4}
Assuming that the assumptions of the Theorem \ref{theo2} hold.
Then, for any $\gamma>0$ we have
\begin{equation}
\displaystyle \limsup_{n\longrightarrow \infty} \displaystyle \sup_{h_{n}^{\prime}\leq h\leq h_{n}^{\prime\prime}} \frac{\sqrt{(nh)^{\alpha}}
\left|\widehat{{\cal D}}_{\alpha}^R(\widehat{f}_{n,h},g)-{\cal D}_{\alpha}^R(f,g)\right|
 }{\sqrt{(\log(1/h)\vee \log \log n)^{\alpha}}}\leq \frac{1}{(1-\alpha)\gamma^\alpha}\zeta(\mathrm{I}) \ \ \hbox{ a.s }, \nonumber
\end{equation}  
\end{Coro}
\vskip1ex  
\noindent
The proof of Corollaries \ref{coro3} and \ref{coro4} are given in Section \ref{proof}.
\vskip1ex  
\noindent
Using the techniques developed in \cite{r9} , the Corollaries (\ref{coro3}) and (\ref{coro4}) lead to the
construction of asymptotic $100\%$ certainty intervals for the true divergences ${\cal D}_{\alpha}^T(f,g)$, ${\cal D}_{\alpha}^R(f,g)$.
Now, assume that there exists a sequence $\{\mathrm{I}_n \}_{n\geq1}$ of strictly nondecreasing compact subsets of $\mathrm{I}$, such that
$\mathrm{I}=\cup_{n\geq1}\mathrm{I}_n.$ For the estimation of the support $\mathrm{I}$ we may refer to (\cite{r12}) and the references therein. 
Throughout, we let $h \in [h^\prime_n , h^{\prime\prime}_n]$, where $h^\prime_n$ and $h^{\prime\prime}_n$
are as in Corollaries (\ref{coro3}) and (\ref{coro4}). Chose an estimator of $\zeta(\mathrm{I})$ in the Corollaries (\ref{coro3}) and (\ref{coro4}) as the form
$$\zeta_n(\mathrm{I}_n)=\sup_{x\in \mathrm{I}_n}\left\{\widehat{f}_{n,h}(x)\int_{\mathrm{R}^d} K^2(u)du\right\}^{\alpha/2}.$$
Thus, we have 
$$\mathrm{P}\left(\left|\zeta_n(\mathrm{I}_n)/\zeta(\mathrm{I})-1\right|\geq\varepsilon\right)\rightarrow 0,\ \ \hbox{ as }\ \ n\rightarrow \infty\ \ \hbox{ for each }\ \ 
\varepsilon>0.$$
Consequently, by defining the quantities
\begin{eqnarray}
 \mathrm{B}_{n}^T&=& \frac{1}{1-\alpha}\zeta_n(\mathrm{I}_n)\int_{\mathrm{R}^d}g^{1-\alpha}dx\times\sqrt{\left(\frac{\log(1/h)\vee \log \log n}{nh}\right)^{\alpha}},\\ \nonumber
\mathrm{B}_n^R&=& \frac{1}{\gamma^\alpha(1-\alpha)}\zeta_n(\mathrm{I}_n)\times\sqrt{\left(\frac{\log(1/h)\vee \log \log n}{nh}\right)^{\alpha}} \nonumber
\end{eqnarray}
we get from Corollaries (\ref{coro3}) and (\ref{coro4}),
$$\mathrm P\left(\frac{1}{B_n^T}\left|\widehat{{\cal D}}_{\alpha}^T(\widehat{f}_{n,h},g)-{\cal D}_{\alpha}^T(f,g)\right|>1+\varepsilon\right)\longrightarrow0,\ \ \ n\longrightarrow\infty.$$
and
$$\mathrm P\left(\frac{1}{B_n^R}\left|\widehat{{\cal D}}_{\alpha}^R(\widehat{f}_{n,h},g)-{\cal D}_{\alpha}^R(f,g)\right|>1+\varepsilon\right)\longrightarrow0,\ \ \ n\longrightarrow\infty.$$
Thus, we obtain asymptotic certainty interval for ${\cal D}_{\alpha}^T(f,g)$ and ${\cal D}_{\alpha}^R(f,g)$ in the following sense.\\
For each $0 < \varepsilon < 1$, we have, as $n\rightarrow\infty$,
$$\mathrm P\left({\cal D}_{\alpha}^R(f,g)\in \left[\widehat{{\cal D}}_{\alpha}^T(\widehat{f}_{n,h},g)-\frac{1}{B_n^T}(1+\varepsilon),
\widehat{{\cal D}}_{\alpha}^T(\widehat{f}_{n,h},g)+\frac{1}{B_n^T}(1+\varepsilon)\right]\right)\longrightarrow1.
$$
and
$$\mathrm P\left({\cal D}_{\alpha}^R(f,g)\in \left[\widehat{{\cal D}}_{\alpha}^R(\widehat{f}_{n,h},g)-\frac{1}{B_n^R}(1+\varepsilon),
\widehat{{\cal D}}_{\alpha}^R(\widehat{f}_{n,h},g)+\frac{1}{B_n^R}(1+\varepsilon)\right]\right)\longrightarrow1.
$$
Finally, we will say that the intervals
$$ \left[\widehat{{\cal D}}_{\alpha}^T(\widehat{f}_{n,h},g)-\frac{1}{B_n^T},
\widehat{{\cal D}}_{\alpha}^T(\widehat{f}_{n,h},g)+\frac{1}{B_n^T}\right],$$
and
$$ \left[\widehat{{\cal D}}_{\alpha}^R(\widehat{f}_{n,h},g)-\frac{1}{B_n^R},
\widehat{{\cal D}}_{\alpha}^R(\widehat{f}_{n,h},g)+\frac{1}{B_n^R}\right],$$
provide  asymptotic $100\%$ certainty intervals for the divergences ${\cal D}_{\alpha}^T(f,g)$ and ${\cal D}_{\alpha}^R(f,g)$.

\section{Concluding remarks and future works}
We have addressed the problem of nonparametric estimation of a class of divergence measures.
We are focusing on the R\'enyi-$\alpha$ and the Tsallis-$\alpha$ divergence measures. 
Under our study, one can easily deduced Kullback-Leibler, Hellinger and Bhattacharyya nonparmetric estimators.
The results presented in this work are general, since the required conditions are fulfilled by a large class of densities.
We mention that the estimator $\widehat{{\cal D}}_{\alpha}(\widehat{f}_{n,h_n},g)$ in (\ref{divestim}) can be calculated by using a Monte-Carlo method under the density $g$. 
And a pratical choice of $\gamma_n$ is $\beta(\log n)^\delta$ where $\beta>0$ and $\delta\geq0$. \\
It will be interesting to enrich our results presented here by an additional
uniformity in term of $\gamma_n$ in the supremum appearing in all our theorems, which
requires non trivial mathematics, this would go well beyond the scope of the
present paper. Another direction of research is to obtain results, in the case where the continuous distributions $f$ and $g$ are both unknown.

\section{Proofs of main results}
\label{proof}
\textbf{Proof of Lemma \ref{lem1}.}
To prove the strong consistency of $\widehat{{\cal D}}_{\alpha}(\widehat{f}_{n,h_n},g)$, we use the following expression
$$ \mathrm{\widehat{E}}\widehat{{\cal D}}_{\alpha}(\widehat{f}_{n,h_n},g):= \int_{\textit{A}_{n,h_n}}\left(\mathrm{E}\widehat{f}_{n,h_n}(x)\right)^\alpha
g^{1-\alpha}(x)dx,$$
where $ \textit{A}_{n,h_n}=\lbrace x\in\mathrm{R}^d, \ \  \widehat{f}_{n,h_n}(x) \geq \gamma_n\rbrace $ and $\gamma_n\downarrow0$ is a sequence of positive constant.
Define
$$\Delta_{n,1,h_n}:=\underbrace{\widehat{{\cal D}}_{\alpha}(\widehat{f}_{n,h_n},g)-\mathrm{\widehat{E}}\widehat{{\cal D}}_{\alpha}(\widehat{f}_{n,h_n},g)}.$$
We have
\begin{eqnarray}
|\Delta_{n,1,h_n}|
&=& \left|\int_{\textit{A}_{n,h_n}}\left( \widehat{f}_{n,h_n}^{\alpha}(x)-\left(\mathrm{E}\widehat{f}_{n,h_n}(x)\right)^{\alpha}\right)g^{1-\alpha}(x)dx\right|\\ \nonumber
&&\leq \int_{\textit{A}_{n,h_n}}\left|\widehat{f}_{n,h_n}^{\alpha}(x)-\left(\mathrm{E}\widehat{f}_{n,h_n}(x)\right)^{\alpha}\right|g^{1-\alpha}(x)dx\\ \nonumber
&&\leq\sup_{x\in\mathrm{R}^d}\left|\widehat{f}_{n,h_n}^{\alpha}(x)-\left(\mathrm{E}\widehat{f}_{n,h_n}(x)\right)^{\alpha}\right|
\int_{\textit{A}_{n,h_n}}g^{1-\alpha}(x)dx.\nonumber
\end{eqnarray}
\noindent Since $h(x)=x$ is a 1-Lipschitz function, for $0< \alpha< 1 $ then
$ \mid (h(x))^{\alpha}-(h(y))^{\alpha} \mid \leq \mid h(x)-h(y)\mid ^{\alpha} $.\\
Therefore for $0< \alpha<1$, we have 
$$
\left|\widehat{f}_{n,h_n}^{\alpha}(x)-\left(\mathrm{E}\widehat{f}_{n,h_n}(x)\right)^{\alpha}\right|\leq \left|\widehat{f}_{n,h_n}(x)- \mathrm{E}\widehat{f}_{n,h_n}(x)\right|^{\alpha}
\leq\left|\left|\widehat{f}_{n,h_n}- \mathrm{E}\widehat{f}_{n,h_n}\right|\right|^{\alpha}_\infty,
$$
where $\left\|\cdot \right\|_{ \infty }$ denotes, as usual, the supremum norm, i.e., 
$\left\|\varphi\right\|_{ \infty }:=\sup_{x\in\mathrm R}|\varphi(x)|$.
Hence, 
\begin{equation}
 \label{result1}
|\Delta_{n,1,h_n}|\leq \left|\left|\widehat{f}_{n,h_n}- \mathrm{E}\widehat{f}_{n,h_n}\right|\right|^{\alpha}_\infty \int_{\textit{A}_{n,h_n}}g^{1-\alpha}(x)dx. 
\end{equation}
Finaly,
\begin{equation}
 \label{terme1}
|\Delta_{n,1,h_n}|
\leq\left|\left|\widehat{f}_{n,h_n}- \mathrm{E}\widehat{f}_{n,h_n}\right|\right|^{\alpha}_\infty\int_{\mathrm{R}^d}g^{1-\alpha}(x)dx.
\end{equation}
\vskip1ex  
\noindent
We now impose some slightly more general assumptions on the kernel $K(\cdot)$ than
that of Theorem \ref{theo1}. Consider the class of functions
$$ \mathcal K:= \left\{ K\left((x-\cdot)/h\right): h>0, \ x\in \mathrm{R}^d \right\}.$$
For $\varepsilon >0$, set $N(\varepsilon , {\cal K })= \sup _{Q}N(\kappa \varepsilon , {\cal K}, d_{Q})$,  where the supremum is taken
over all probability measures $Q$ on $(\mathrm{R}^{d}, {\cal B } )$, where ${\cal B } $ represents the $\sigma$-field of Borel sets of
$\mathrm{R}^d$. Here, $d_{Q}$ denotes the $L_{2}(Q)$-metric and $N( \kappa\varepsilon , {\cal K}, d_{Q})$ is the minimal number of balls 
$\lbrace \psi: d_{Q}(\psi,\psi^{\prime})<\varepsilon \rbrace$ of $d_{Q}$-raduis $\varepsilon$ needed to cover ${\cal K}$. \\
We assume that ${\cal K}$ satisfies the following uniform entropy condition.
\vskip1ex  

(\textbf{K.6}) for some $C>0$ and $\nu >0$, $ N(\varepsilon ,{\cal K})\leq C\varepsilon ^{-\nu}, 0<\varepsilon <1. $
\vskip1ex  

(\textbf{K.7}) ${\cal K } $ is a pointwise measurable class, that is there exists a countable sub-class ${\cal K }_{0} $ of ${\cal K } $ such 
 
that we can find for any function $\psi\in {\cal K }$ a  sequence of functions $\lbrace \psi_{m}: m\geq 1\rbrace$ in ${\cal K}_{0}$ for
which 
$$ \psi_{m}(z)\longrightarrow \psi(z), \quad z\in \mathrm{R}^d.$$
This condition is discussed in \cite{r33} . It is satisfied whenever $K$ is right
continuous.
\vskip1ex  
\noindent \textit{\small Remark that condition (\textbf{K.6}) is satisfied whenever (\textbf{K.1}) holds,
i.e., $K(\cdot)$ is of bounded variation on $\mathrm{R}^d$ (in the sense of Hardy and Kauser, see,
e.g. \cite{r4,r35} and \cite{r20}. Condition (\textbf{K.7}) is satisfied whenever (\textbf{K.2}) holds, \textit{i.e.}, $K(\cdot)$
is right continuous (refer to \cite{r9,r15}and the references therein).}
\vskip1ex
\noindent 
From Theorem 1 in \cite{r15}, whenever $K(\cdot)$ is measurable and satisfies (\textbf{K.3-4-6-7}), and when $f(\cdot)$ is bounded, we have
for each pair of sequence $(h_{n}^{\prime})_{n\geq1}$, $(h_{n}^{\prime\prime})_{n\geq1}$ such that 
$0< h_{n}^{\prime} < h \leq h_{n}^{\prime\prime}\leq1$, together with $h_{n}^{\prime\prime} \rightarrow 0 $ and
$\displaystyle nh_{n}^{\prime}/\log(n)\rightarrow \infty $ as $n \longrightarrow \infty $, with probability 1
\begin{equation}
 \label{mason2005}
 \displaystyle\sup_{h_{n}^{'}\leq h \leq h_{n}^{''}}\left|\left|\widehat{f}_{n,h}- \mathrm{E}\widehat{f}_{n,h}\right|\right|_\infty =
O\left(\sqrt{\frac{\log(1/h_n^\prime)\vee \log \log n}{nh_{n}^\prime}}\right). 
\end{equation}
Since $\int_{\mathrm{R}^d}g^{1-\alpha}(x)dx<\infty$, in view of (\ref{terme1}) and (\ref{mason2005}),  we obtain with probability 1
\begin{equation}
\label{result11}
\displaystyle\sup_{h_{n}^{'}\leq h \leq h_{n}^{''}}\left|\Delta_{n,1,h}\right| =
O\left(\left(\frac{\log(1/h_n^\prime)\vee \log \log n}{nh_{n}^\prime}\right)^{\alpha/2}\right).
\end{equation}
It concludes the proof of the lemma.

\noindent  \textbf{Proof of Lemma \ref{lem2}.}\\
Let $A_{n,h_n}^c$ be the complement of $A_{n,h_n}$ in $\mathrm{R}^d$ (\textit{i.e}, $A_{n,h_n}^c=\{x\in\mathrm{R}^d,\ \widehat{f}_{n,h_n}<\gamma_n\}$).
We have
\begin{eqnarray}
\mathrm{\widehat{E}}\widehat{{\cal D}}_{\alpha}(\widehat{f}_{n,h_n},g)
-{\cal D}_{\alpha}(f,g)&=&\Delta_{n,2,h_n}+\Delta_{n,3,h_n},\nonumber
\end{eqnarray}
with
$$\Delta_{n,2,h_n}:=\int_{\textit{A}_{n,h_n}}\left(\left(\mathrm{E}\widehat{f}_{n,h_n}(x)\right)^\alpha-f^\alpha(x)\right)
g^{1-\alpha}(x)dx$$
and
$$\Delta_{n,3,h_n}:=\int_{\textit{A}_{n,h_n}^c}f^\alpha(x)g^{1-\alpha}(x)dx.\ \ \ \ \ \ \ $$
\underline{Term $\Delta_{n,2,h_n}$}.
Repeat the arguments above in the terms $\Delta_{n,1,h_n}$ with the formal change of $\widehat{f}_{n,h_n}$ by $f$. 
We show that, for any $n\geq1$,
\begin{equation}
 \label{result2}
|\Delta_{n,2,h_n}| \leq\left|\left|\mathrm{E}\widehat{f}_{n,h_n}-f\right|\right|^{\alpha}_\infty\int_{\hbox{ A }_{n,h_n}}g^{1-\alpha}(x)dx,
\end{equation}
which implies
\begin{equation}
 \label{term2}
|\Delta_{n,2,h_n}| \leq\left|\left|\mathrm{E}\widehat{f}_{n,h_n}-f\right|\right|^{\alpha}_\infty\int_{\mathrm{R}^d}g^{1-\alpha}(x)dx.
\end{equation}
On the other hand, we know (see, e.g,\cite{r15} ), that since the density $f(\cdot)$ is
uniformly Lipschitz and continuous, we have for each sequences  $h_n^\prime<h< h_n^{\prime\prime}<1$, with  $h_n^{\prime\prime}\rightarrow0$,  as $n\rightarrow\infty$,
\begin{equation}
 \label{mason20051}
\sup_{h_{n}^{'}\leq h \leq h_{n}^{''}}\left|\left|\mathrm{E}\widehat{f}_{n,h_n}-f\right|\right|_\infty=O(h_n^{\prime\prime {1/d}}).
\end{equation}
Thus,
\begin{equation}
\label{form1}
 \displaystyle\sup_{h_{n}^{'}\leq h \leq h_{n}^{''}}\left|\Delta_{n,2,h}\right| =
O(h_n^{\prime\prime {\alpha/d}}). 
\end{equation}
\underline{Term $\Delta_{n,3,h_n}$}.
It is obsious to see that
\begin{eqnarray}
\left|\Delta_{n,3,h_n}\right|&=&\int_{\textit{A}_{n,h_n}^c} |f^{\alpha}(x)|g^{1-\alpha}(x)dx\\ \nonumber
 &&\leq \int_{\hbox{ A }_{n,h_n}^c}\left|\mathrm{E}\widehat{f}_{n,h_n}(x)-f^{\alpha}(x)\right|g^{1-\alpha}(x)dx
+\int_{\hbox{ A }_{n,h_n}^c}\mathrm{E}\widehat{f}_{n,h_n}(x)g^{1-\alpha}(x)dx\\ \nonumber
&&\leq\left|\left|\mathrm{E}\widehat{f}_{n,h_n}-f\right|\right|^{\alpha}_\infty\int_{\hbox{ A }_{n,h_n}^c}g^{1-\alpha}(x)dx
+\gamma_n^\alpha\int_{\hbox{ A}_{n,h_n}^c}g^{1-\alpha}(x)dx.
\end{eqnarray}
Thus,
\begin{eqnarray}
\label{esult3} 
\left|\Delta_{n,3,h_n}\right| & \leq & \left(\left|\left|\mathrm{E}\widehat{f}_{n,h_n}-f\right|\right|^{\alpha}_\infty+
\gamma_n^\alpha\right)\int_{\hbox{ A }_{n,h_n}^c}g^{1-\alpha}(x)dx.
\end{eqnarray}
Hence,
\begin{eqnarray}
\label{term2} 
\left|\Delta_{n,3,h_n}\right| & \leq & \left(\left|\left|\mathrm{E}\widehat{f}_{n,h_n}-f\right|\right|^{\alpha}_\infty+
\gamma_n^\alpha\right)\int_{\mathrm{R}^d}g^{1-\alpha}(x)dx.
\end{eqnarray}
Thus, in view of (\ref{mason20051}), we get
\begin{eqnarray}
\label{form2}
 \displaystyle\sup_{h_{n}^{'}\leq h \leq h_{n}^{''}}\left|\Delta_{n,3,h_n}\right| &=& 
O\left(\gamma_n^\alpha\vee h_n^{\prime\prime {\alpha/d}}\right)
\end{eqnarray}
Finaly, in view of (\ref{form1}) and (\ref{form2}), we get  
\begin{equation}
\label{bias}
 \displaystyle\sup_{h_{n}^{'}\leq h \leq h_{n}^{''}}\left|\mathrm{\widehat{E}}\widehat{{\cal D}}_{\alpha}(\widehat{f}_{n,h},g)
-{\cal D}_{\alpha}(f,g)\right|=O\left(\gamma_n^\alpha\vee h_n^{\prime\prime {\alpha/d}}\right).
\end{equation}
It concludes the proof of the lemma.

\noindent  \textbf{Proof of Theorem \ref{theo1}.} We have
\begin{eqnarray}
\left|\widehat{{\cal D}}_{\alpha}^R(\widehat{f}_{n,h_n},g)-{\cal D}_{\alpha}^R(f,g)\right|
&\leq & \left|\widehat{{\cal D}}_{\alpha}(\widehat{f}_{n,h},g)
-\mathrm{\widehat{E}}\widehat{{\cal D}}_{\alpha}(\widehat{f}_{n,h},g)\right|+\left|\mathrm{\widehat{E}}\widehat{{\cal D}}_{\alpha}(\widehat{f}_{n,h},g)
-{\cal D}_{\alpha}(f,g)\right|.\nonumber
\end{eqnarray}
Combinating the Lemmas (\ref{lem1}) and (\ref{lem2}), we obtain
\begin{eqnarray}
\displaystyle\sup_{h_{n}^{'}\leq h \leq h_{n}^{''}} \left|\widehat{{\cal D}}_{\alpha}(\widehat{f}_{n,h_n},g)-{\cal D}_{\alpha}(f,g)\right|
&=& O\left(\left(\frac{\log(1/h_n^\prime)\vee \log \log n}{nh_{n}^\prime}\right )^{\alpha/2} \right)+O\left(\gamma_n^\alpha\vee h_n^{\prime\prime {\alpha/d}}\right).\nonumber 
\end{eqnarray}
It concludes the proof of the Theorem.

\noindent \textbf{Proof of Corollary \ref{coro1}.} Remark that
\begin{eqnarray}
\widehat{{\cal D}}_{\alpha}^T(\widehat{f}_{n,h_n},g)-{\cal D}_{\alpha}^T(f,g)
&=& \frac{1}{\alpha-1}\left(\widehat{{\cal D}}_{\alpha}(\widehat{f}_{n,h_n},g)-{\cal D}_{\alpha}(f,g)
\right). \nonumber 
\end{eqnarray}
Using the Theorem (\ref{theo1}), we have 
\begin{eqnarray}
 \displaystyle\sup_{h_{n}^{'}\leq h \leq h_{n}^{''}}\left|\widehat{{\cal D}}_{\alpha}^T(\widehat{f}_{n,h_n},g)-{\cal D}_{\alpha}^T(f,g)\right| &=&
O\left(\left(\frac{\log(1/h_n^\prime)\vee \log \log n}{nh_{n}^\prime}\right)^{\alpha/2}\vee\gamma_n^\alpha\vee h_n^{\prime\prime {\alpha/d}}\right),\nonumber
\end{eqnarray}
and the Corollary \ref{coro1} holds
 
\noindent \textbf{Proof of Corollary \ref{coro2}.}
A first order taylor expansion of $y\mapsto\log y$ arround $y=y_0>0$ and $y=\widehat{y}>0$ gives
\begin{equation}
 \log \widehat{y}=\log y_0+\frac{1}{y_0}(\widehat{y}-y_0)+o(||\widehat{y}-y_0||).\nonumber
\end{equation}
Remark that from Theorem \ref{theo1},
\begin{equation}
 \displaystyle\sup_{h_{n}^{'}\leq h \leq h_{n}^{''}}\left|\widehat{{\cal D}}_{\alpha}(\widehat{f}_{n,h},g)-{\cal D}_\alpha(f,g)\right| =
O\left(\left(\frac{\log(1/h_n^\prime)\vee \log \log n}{nh_{n}^\prime}\right)^{\alpha/2}\vee\gamma_n^\alpha\vee h_n^{\prime\prime {\alpha/d}}\right),\nonumber
\end{equation}
which turn, implies that
\begin{equation}
 \displaystyle \lim _{n\rightarrow \infty }\sup_{h_{n}^{\prime}\leq h\leq h_{n}^{\prime\prime}}
\left|\widehat{{\cal D}}_{\alpha}(\widehat{f}_{n,h},g)-{\cal D}_{\alpha}(f,g)\right|=0 \quad \quad a.s. \nonumber
\end{equation}
Thus, for all 
\begin{eqnarray}
\widehat{{\cal D}}_{\alpha}^R(\widehat{f}_{n,h_n},g)-{\cal D}_{\alpha}^R(f,g) \nonumber
&=&\frac{1}{\alpha-1}
\left(\log \widehat{{\cal D}}_{\alpha}(\widehat{f}_{n,h},g)-\log {\cal D}_{\alpha}(f,g)\right)\\ \nonumber
&=&\frac{1}{(\alpha-1){\cal D}_{\alpha}(f,g)}
\left(\widehat{{\cal D}}_{\alpha}
(\widehat{f}_{n,h},g)-
{\cal D}_{\alpha}(f,g)\right)
 \\ \nonumber
& &\ \ \ \ \ \ \ \ \ \ +o\left( \left|\left|\widehat{{\cal D}}_{\alpha}(\widehat{f}_{n,h},g)-{\cal D}_{\alpha}(f,g)
\right|\right|
 \right). \nonumber
\end{eqnarray}

Consequently
\begin{equation}
 \displaystyle\sup_{h_{n}^{'}\leq h \leq h_{n}^{''}}\left|\widehat{{\cal D}}_{\alpha}^R(\widehat{f}_{n,h_n},g)-{\cal D}_{\alpha}^R(f,g)\right| =
O\left(\left(\frac{\log(1/h_n^\prime)\vee \log \log n}{nh_{n}^\prime}\right)^{\alpha/2}\!\!\vee\!\!\gamma_n^\alpha\vee h_n^{\prime\prime {\alpha/d}}\right),\nonumber
\end{equation}
and the Corollary \ref{coro2} holds.
 
\noindent \textbf{Proof of Theorem \ref{theo2}.}
Under conditions $(\textbf{F.1})$, $(\textbf{K.5})$ and using Taylor expansion of order $s$ we get, for $x \in \mathrm{I} $,
$$ \vert \mathrm{E} \widehat{f}_{n,h_{n}}-f(x) \vert = \frac{h^{s/d}}{s!}\left\vert \int \displaystyle \sum_{k_{1}+...+k_{d}} t_{1}^{k_{1}}...t_{d}^{k_{d}} \frac{ \partial^{s} f(x-h\theta t) }{\partial x_{1}^{k_{1}} ... \partial x_{1}^{k_{d}}} K(t)dt\right\vert  $$ 
where $ \theta = (\theta_{1}, ..., \theta_{d} ) $ and $ 0< \theta_{i} <1,\ \  i=1,...;d $ Thus a straightforward application of Lebesgue dominated convergence theorem gives, for $n$ large enough,
 $$ \displaystyle \sup_{x \in \mathrm{I} } \vert \mathrm{E}\widehat{f}_{n,h}(x)- f(x) \vert = O(h_{n}^{''}) $$
 Let $\mathrm{J}$ be a nonempty compact subset of the interior of 
 $\mathrm{I}$ 
 (say $\displaystyle 
 \mathring{I})$). \\
First, note that we have from Corollary 3.1.2. p. 62 of Viallon (2006) (see also, \cite{r3} , statement (4.16)).
\begin{equation}
\label{valliron}
 \displaystyle \limsup_{n\rightarrow \infty } \displaystyle \sup_{h_{n}^{\prime}\leq h\leq h_{n}^{\prime}} \displaystyle \sup_{x\in \mathrm{J}}
 \frac{\sqrt{nh}\vert \widehat{f}_{n,h}(x)-f(x) \vert}{\sqrt{\log (1/h)\vee \log\log n }}= \displaystyle \sup_{x \in \mathrm{J} }\left( f(x)\int_{\mathrm{R}^{d}} K^{2}(t)dt\right)^{1/2}
\end{equation}
Set, for all $n\geq 1$,
\begin{eqnarray}
\nonumber\pi_{n}(\mathrm{J}) & =&  \left\vert \int_{\mathrm{J}} \left( \widehat{f}_{n,h}^{\alpha}(x)- f^{\alpha}(x)\right)g^{1-\alpha}(x) dx \right\vert \\ 
\nonumber& \leq &  \int_{\mathrm{J}} \vert \widehat{f}_{n,h}^{\alpha}(x)- f^{\alpha}(x) \left \vert g^{1-\alpha}(x) dx \right) \\ 
\nonumber &\leq & \int_{\mathrm{J}} \vert  \widehat{f}_{n,h}(x)- f(x) \vert^{\alpha}  g^{1-\alpha}(x) dx\ \ \ \ \hbox{ since }\ \ \alpha\in]0,1[,
\\ 
\label{valiron2d}
& \leq &\displaystyle \sup_{x\in \mathrm{J}} \vert \widehat{f}_{n}(x)- f(x) \vert^{\alpha} \int_{\mathrm{J}} g^{1-\alpha}(x)dx,\\
\label{valiron2}
\displaystyle & \leq & \sup_{x\in \mathrm{J}} \vert \widehat{f}_{n}(x)- f(x) \vert^{\alpha} \int_{\mathrm{R}^d} g^{1-\alpha}(x)dx.
\end{eqnarray}
One fined, by combining (\ref{valliron}) and (\ref{valiron2})
\begin{equation}
\displaystyle \limsup_{n\rightarrow \infty } \displaystyle \sup_{h_{n}^{\prime}\leq h\leq h_{n}^{\prime\prime}} 
\frac{\sqrt{(nh)^{\alpha}}\pi_{n}(\mathrm{J} ) }{\sqrt{(\log (1/h)\vee \log\log n )^{\alpha}}}\nonumber
\end{equation}
\begin{equation}
 \label{dem}
 ~~~~~~~~~~~~~~~~~~~~~~~~~~~~~~~~~~~~~~~~~~~~\leq 
 \displaystyle \sup_{x \in \mathrm{J} }
 \left \lbrace 
 \left( f(x) \int_{\mathrm{R}^{d}} K^{2}(t)dt 
 \right)^{\alpha/2}
  \right \rbrace \int_{\mathrm{R}^d}g^{1-\alpha}(x)dx.
\end{equation}
Let $\lbrace \mathrm{J}_{\ell}\rbrace,\ \  \ell=1,2,..., $
be a sequence of nondecreasing nonempty compact subsets of $\displaystyle \mathring{I} $ such that $$ \displaystyle \bigcup_{\ell\geq 1}\mathrm{J}_{\ell}= \mathring{I} $$
Now, from (\ref{dem}), it is straightforward to observe that
\begin{eqnarray}
\displaystyle \limsup_{\ell\rightarrow \infty } 
&&\limsup_{n\rightarrow \infty }
 \displaystyle \sup_{h_{n}^{\prime}\leq h\leq h_{n}^{\prime\prime}} 
\frac{\sqrt{(nh)^{\alpha}}\pi_{n}(\mathrm{J}_{\ell} ) }{\sqrt{(\log (1/h)\vee \log\log n )^{\alpha}}} \nonumber \\
&& \leq \displaystyle \limsup_{\ell\rightarrow \infty } \displaystyle \sup_{x \in \mathrm{J}_{\ell} }\left\lbrace  f(x)\int_{\mathrm{R}^{d}} K^{2}(t)dt \rbrace^{\alpha/2}
\right\rbrace \int_{\mathrm{R}^d}  g^{1-\alpha}(x)dx \nonumber \\
 && \leq  \displaystyle \sup_{x \in \mathrm{I} }\left\lbrace \left(  f(x)\int_{\mathrm{R}^{d}} K^{2}(t)dt \right)^{\alpha/2} \right\rbrace
\int_{\mathrm{R}^d}  g^{1-\alpha}(x)dx  \nonumber
\end{eqnarray}
The proof of Theorem \ref{theo2} is completed.

\noindent \textbf{Proof of Corollary \ref{coro3}.}
A direct application of the Theorem \ref{theo2} leeds to the Corollary \ref{coro3}. 
 
\noindent \textbf{Proof of Corollary \ref{coro4}.}
Here again, 
set, for all $n\geq 1$,
\begin{eqnarray}
\label{valiron22}
\nonumber\eta_{n}(\mathrm{J})=  \left\vert \frac{1}{\alpha-1}\left(\log\int_{\mathrm{J}}\widehat{f}_{n,h}^{\alpha}(x)g^{1-\alpha}-\log\int_{\mathrm{J}}f^{\alpha}(x)g^{1-\alpha}(x) dx \right)\right\vert. 
\end{eqnarray}
A first order Taylor expansion of $\log (y)$ leeds to
\begin{eqnarray}
\eta_{n}(\mathrm{J})&&\leq\frac{1}{1-\alpha}\frac{1}{\int_{\mathrm{J} } f^{\alpha}(x) g^{1-\alpha}(x)dx }
\left|\int_{\mathrm{J}} \left( \widehat{f}_{n,h}^{\alpha}(x)- f^{\alpha}(x) \right) g^{1-\alpha}(x)dx\right|+ o\left(||\widehat{f}_{n,h}^{\alpha}-f||^\alpha_\infty\right), \nonumber \\
&&\leq\frac{1}{1-\alpha}\frac{1}{\int_{\mathrm{J} } f^{\alpha}(x) g^{1-\alpha}(x)dx }
\pi_n(\mathrm{J})+ o\left(||\widehat{f}_{n,h}^{\alpha}-f||^\alpha_\infty\right), \nonumber
\end{eqnarray}
Using condition $(F.1)$, $f(\cdot)$ is compactly supported), $f(\cdot)$ is bounded away from
zero on its support, thus, we have for $n$ enough large, there exists $\gamma > 0$, such
that $f(x)> \gamma $, for all $x$ in the support of $f(\cdot)$.
From (\ref{valiron2d}), we have
$$\pi_{n}(\mathrm{J} ) \leq \displaystyle \sup_{x\in \mathrm{J}} \vert \widehat{f}_{n,h}(x)- f(x) \vert^{\alpha} \int_{\mathrm{J} } g^{1-\alpha}(x)dx.$$
Hence,
\begin{eqnarray}
\eta_{n}(\mathrm{J})&&\leq \frac{1}{1-\alpha}\frac{1}{\gamma^{\alpha}}\frac{1}{\int_{\mathrm{J} } g^{1-\alpha}(x)dx } \displaystyle \sup_{x\in \mathrm{J}}
\vert \widehat{f}_{n}(x)- f(x) \vert^{\alpha} \int_{\mathrm{J^{d}}} g^{1-\alpha}(x)dx  \nonumber \\
 && \leq \frac{1}{1-\alpha}\frac{1}{\gamma^{\alpha}}\displaystyle \sup_{x\in \mathrm{J}} \vert \widehat{f}_{n}(x)- f(x) \vert^{\alpha} \nonumber
\end{eqnarray}
One fined, by combining the last equation with (\ref{valliron})
$$ \displaystyle \limsup_{n\rightarrow \infty } \displaystyle \sup_{h_{n}^{\prime}\leq h\leq h_{n}^{\prime\prime}}
\frac{\sqrt{(nh)^{\alpha}}\eta _{n}(\mathrm{J} ) }{\sqrt{(\log (\frac{1}{h})\vee \log\log n )^{\alpha}}} 
\leq \frac{1}{1-\alpha}\frac{1}{\gamma^{\alpha}} \displaystyle \sup_{x \in \mathrm{J} }\left\lbrace \left( f(x)\int_{\mathrm{R}^{d}} K^{2}(t)dt \right)^{\alpha/2}\right\rbrace 
$$
\begin{eqnarray}
\displaystyle \limsup_{ell\rightarrow \infty } \displaystyle \limsup_{n\rightarrow \infty } \displaystyle \sup_{h_{n}^{\prime}\leq h\leq h_{n}^{\prime\prime}}
  \frac{\sqrt{(nh)^{\alpha}}\eta _{n}(\mathrm{J_{\ell}} )}{\sqrt{(\log (1/h)\vee \log\log n )^{\alpha}}}~~~~~~~~~~~~~~~~~~~~~~~~~~~
\end{eqnarray}
\begin{eqnarray}
 & &
\leq \frac{1}{1-\alpha}\frac{1}{\gamma^{\alpha}} \displaystyle \limsup_{l\rightarrow \infty } \displaystyle \sup_{x \in \mathrm{J}_{\ell
} }\left\lbrace  f(x)\int_{\mathrm{R}^{d}} K^{2}(t)dt \rbrace^{\alpha/2}\right\rbrace \\  \nonumber
 && \leq  \frac{1}{1-\alpha}\frac{1}{\gamma^{\alpha}}  \displaystyle \sup_{x \in \mathrm{I} }\left\lbrace \left(  f(x)\int_{\mathrm{R}^{d}} K^{2}(t)dt \right)^{\alpha/2}\right\rbrace  \nonumber
\end{eqnarray}
The proof of Corollary is completed.
\section*{Acknowledgements}

\end{document}